\documentstyle[12pt,psfig]{article}
\newcommand{\EQ}{\begin{equation}}
\newcommand{\EN}{\end{equation}}
\newcommand{\bea}{\begin{eqnarray}}
\newcommand{\eea}{\end{eqnarray}}

\newcommand{\th}{\theta}

\begin{document}
\topmargin 0pt
\oddsidemargin 5mm
\renewcommand{\thefootnote}{\arabic{footnote}}
\newpage
\setcounter{page}{0}
\begin{titlepage}
\hspace{11.8cm}SISSA-3/2001/FM
\vspace{1cm}
\vspace{0.5cm}
\begin{center}
{\large {\bf One-point functions in integrable quantum field theory
at finite temperature}}\\
\vspace{1.8cm}
{\large G. Delfino}\footnote{E-mail: delfino@sissa.it} \\ \vspace{0.5cm}
{\em International School for Advanced Studies and Istituto Nazionale
di Fisica Nucleare 34014 Trieste, Italy}
\end{center}
\vspace{1.2cm}

\renewcommand{\thefootnote}{\arabic{footnote}}
\setcounter{footnote}{0}

\begin{abstract}
\noindent
We determine the form factor expansion of the one-point functions in 
integrable quantum field theory at finite temperature and find that 
it is simpler than previously conjectured.
We show that no singularities are left in the final expression provided
that the operator is local with respect to the particles and argue 
that the divergences arising in the non-local case are related to 
the absence of spontaneous symmetry breaking on the cylinder.
As a specific application, we give the first terms of the low temperature
expansion of the one-point functions for the Ising model in a magnetic 
field.
\end{abstract}

\vspace{.3cm}

\end{titlepage}

\newpage
Quantum field theory at finite temperature is a subject of both
theoretical and experimental interest. In the euclidean formulation,
it corresponds to work with an immaginary time compactified on a 
circle whose circumference $R$ coincides with the inverse temperature.
The average of a generic product $X$ of local operators is given by
\EQ
\langle X\rangle_R=\frac{\mbox{Tr}\,Xe^{-RH}}{\mbox{Tr}\,e^{-RH}}\,,
\label{average}
\EN
where $H$ is the Hamiltonian. 

In two dimensions, a non-perturbative study of the properties of 
the finite temperature case should be achievable
exploiting the results of integrable quantum field theory.
In particular, a natural idea is that of approaching the computation
of the correlation functions in a way similar to that successfully 
used in the zero temperature case, and evaluate (\ref{average}) on the 
basis of the multi-particle asymptotic states of the zero temperature 
theory. At the end, the problem should be reduced to summing over the 
matrix elements of the operators between the vacuum and the $n$-particle 
asymptotic states (the form factors) which are known exactly in
integrable theories. 

The one-point functions have a non-trivial temperature dependence and
provide the first test that this form factor method needs to pass. 
Following the analysis of the free fermionic case \cite{LLSS}, an 
extension to the generic integrable case exploiting also the results of the 
thermodynamic Bethe ansatz (TBA) was conjectured in Ref.\,\cite{LM}. 
More precisely, it was proposed that the `pseudoenergy' given by
the TBA should substitute the ordinary energy of the particles in the 
thermal weight function entering the form factor sum. Lukyanov found 
agreement with this conjecture in his semi-classical study of the 
sinh-Gordon model at finite temperature \cite{Lukyanov}.

In this note, we analyse explicitely the form factor expansion of the
one-point functions in finite temperature integrable theories and find
that the final expression, Eq.\,(\ref{final}), is simpler than proposed 
in the past: the weight function is the same than in the free cases and 
all the effects of the interaction are contained in the form factors. 
Along the way we discuss a number of delicate issues originating from the 
fact that the expression (\ref{average}) is a sum over matrix elements 
between identical asymptotic states which contain singularities. We 
explain why and in which cases the final result is free of divergences.

This analysis also allows a clarification of the status of the conjecture
of Ref.\,\cite{LM}. In fact, the expression for the one-point functions
proposed in that work differs from our Eq.\,(\ref{final}) not only in 
the thermal weight, but also in the operator dependent part. In 
Refs.\,\cite{LM,Saleur}, the latter does not coincide with the form 
factors (\ref{parallel}) but is such that its combination 
with the modified weight function reproduces the TBA result for 
the one-point function of the trace of the stress-energy tensor. At present
it is not clear how a similar expression involving the pseudoenergies could 
be derived for other operators. In any case, due to the complicated 
temperature dependence of the pseudoenergies, it would not be the 
explicit low temperature expansion (\ref{final}).

\vspace{.3cm}
Consider an integrable quantum field theory whose spectrum contains a 
single neutral particle of mass $M$. As usual, let us use the rapidity
variable $\th$ to parameterise the energy and momentum of a particle
as $(p^0,p^1)=(M\cosh\th,M\sinh\th)$. Then relativistic invariant 
quantities depend on the rapidity differences only. We denote 
$S(\th_1-\th_2)$ the scattering amplitude of two particles with 
rapidities $\th_1$ and $\th_2$. It satisfies the crossing symmetry
relation
\EQ
S(\th)=S(i\pi-\th)\,\,.
\label{crossing}
\EN
Integrability shows up in two respects \cite{ZZ}. First, it 
forbids any inelastic process, so that the unitarity condition takes 
the simple form
\EQ
S(\th)S(-\th)=1\,\,.
\label{unitarity}
\EN
Second, it induces the complete factorisation of multi-particle scattering 
amplitudes into the product of the two-particle ones. 

Take now a scalar operator $\Phi(x)$ in this theory which is {\em local} with 
respect to the particles, and denote its
matrix elements on the asymptotic multi-particle states as
\EQ
F^\Phi_{m,n}(\th'_m,\ldots,\th'_1|\th_1,\ldots,\th_n)=
\langle\th'_m,\ldots,\th'_1|\Phi(0)|\th_1,\ldots,\th_n\rangle\,\,.
\label{me}
\EN
The contribution of the $n$-particle asymptotic state to 
$\mbox{Tr}\,\Phi(x)\,e^{-HR}$ is
\EQ
f^\Phi_n(R)=\frac{1}{n!}\,\frac{1}{(2\pi)^n}\int d\th_1\ldots d\th_n\,
F^\Phi_{n,n}(\th_n,\ldots,\th_1|\th_1,\ldots,\th_n)\,e^{-E_nR}\,\,,
\label{fn}
\EN
where 
\EQ
E_n=M\sum_{i=1}^n\cosh\th_i\,\,,
\EN
is the total energy of the asymptotic state, and we
used the property $\Phi(x)=e^{iP_\mu x^\mu}\Phi(0)e^{-iP_\mu x^\mu}$.
$f^\Phi_n(R)$ is of order $e^{-nMR}$ for $MR\rightarrow\infty$. 
After defining
\EQ
f^\Phi(R)=\sum_{n=0}^\infty f^\Phi_n(R)\,\,,
\EN
we can write
\EQ
\langle\Phi\rangle_R=\frac{f^\Phi(R)}{f^I(R)}\,\,,
\label{ratio}
\EN
with $I$ denoting the identity operator.

It is our goal to rewrite this rather implicit expression containing 
the matrix elements $F^\Phi_{n,n}$ as a sum over the form factors
$F^\Phi_{0,n}$ which are exactly computable in integrable field theories
\cite{Karowski,Smirnov}. Technically, this problem has several 
points of contact with Ref.\,\cite{nonint}, where non-integrable field
theories were considered as perturbations of the integrable ones. In 
that context the matrix elements $F^\Phi_{m,n}$ determine the first
order correction to the scattering process $m\rightarrow n$ under the 
action of the perturbing operator $\Phi$.

The reduction to the form factors $F^\Phi_{0,n}$ is performed iterating 
$m$ times the crossing relation \cite{Karowski,Smirnov}
\bea
&&F^\Phi_{m,n}(\th'_m,\ldots,\th'_1|\th_1,\ldots,\th_n)= 
F^\Phi_{m-1,n+1}(\th'_m,\ldots,\th'_2|\th'_1+i\pi,\th_1,\ldots,\th_n)+
\nonumber \\
&& 2\pi\sum_{i=1}^{n}\delta(\th'_1-\th_i)\prod_{k=1}^{i-1}
S(\th_k-\th'_1)\,F^\Phi_{m-1,n-1}(\th'_m,\ldots,\th'_2|\th_1,
\ldots,\th_{i-1},\th_{i+1},\ldots,\th_n)\,\,.
\label{cross}
\eea
The second term in the r.h.s. accounts for the disconnected parts
that appear if the crossed particle hits a particle with exactly the
same momentum and annihilates it. It contains the product of the 
scattering amplitudes with the particles that need to be crossed on
the way. We recall in fact that the relation
\EQ
|\ldots\th_i,\th_{i+1}\ldots\rangle=S(\th_i-\th_{i+1})
|\ldots\th_{i+1},\th_{i}\ldots\rangle\,,
\label{exchange}
\EN
is characteristic of integrable field theories. The situation of 
collinearity responsible for the disconnected parts also induces
an `annihilation' pole in the matrix elements with residue \cite{Smirnov}
\EQ
-i\,\mbox{res}_{\th_1=\th_2+i\pi}F^\Phi_{0,n}(|\th_1,\ldots,\th_n)=
\left[1-\prod_{i=1}^nS(\th_i-\th_2)\right]\,
F^\Phi_{0,n-2}(|\th_3,\ldots,\th_n).
\label{res}
\EN
Since (\ref{fn}) contains matrix elements involving two identical sets 
of momenta, it is obvious that we will have to care about both 
annihilation poles and singular disconnected parts containing $\delta(0)$
factors. 
We will see in a moment, however, that these potential sources of trouble 
drop out and leave a final result which is perfectly finite.

Using (\ref{cross}) the first few $f_n^\Phi(R)$ are found to be
\bea
f^\Phi_0 &=& {\cal F}^\Phi_0=\langle 0|\Phi|0\rangle=
\langle\Phi\rangle_{R=\infty}\,,\nonumber \\
f^\Phi_1(R) &=& \frac{1}{2\pi}\,\int d\th_1\,e^{-E_1R}\,[{\cal F}^\Phi_2+
2\pi\delta(0){\cal F}^\Phi_0]\,,\nonumber \\
f^\Phi_2(R) &=& \frac{1}{2}\,\frac{1}{(2\pi)^2}\int d\th_1d\th_2\,
e^{-E_2R}\,[{\cal F}^\Phi_4(\th_1,\th_2)+4\pi(\delta(\th_1-\th_2)S(0)+
\delta(0)){\cal F}^\Phi_2 \nonumber \\
&+& (2\pi)^2(S(0)\delta^2(\th_1-\th_2)+\delta^2(0)){\cal F}^\Phi_0]\,,
\nonumber \\
f^\Phi_3(R) &=&\frac{1}{3!}\,\frac{1}{(2\pi)^3}\int d\th_1d\th_2d\th_3\,
e^{-E_3R}\,[{\cal F}^\Phi_6(\th_1,\th_2,\th_3)+6\pi(2S(0)\delta(\th_1-\th_3)+
\delta(0)){\cal F}^\Phi_4(\th_1,\th_2) \nonumber \\
&+& 3(2\pi)^2(2\delta(\th_1-\th_2)
\delta(\th_1-\th_3)+S(0)\delta^2(\th_1-\th_2)+2S(0)\delta(0)
\delta(\th_1-\th_2)+\delta^2(0)){\cal F}^\Phi_2 \nonumber \\
&+& (2\pi)^3(2\delta(\th_1-\th_2)\delta(\th_1-\th_3)\delta(\th_2-\th_3)+
3S(0)\delta(0)\delta^2(\th_1-\th_2)+\delta^3(0)){\cal F}^\Phi_0]\,,\nonumber
\eea
where we introduced the notation
\EQ
{\cal F}^\Phi_{2n}(\th_1,\ldots,\th_n)=\lim_{\epsilon\rightarrow 0}
F^\Phi_{0,2n}(|\th_n+i\pi+\epsilon,\ldots,\th_1+i\pi+\epsilon,
\th_1,\ldots,\th_n)\,\,.
\label{parallel}
\EN
${\cal F}^\Phi_2(\th_1)$ is a constant by relativistic invariance.
We inserted in Eq.\,(\ref{parallel}) an infinitesimal displacent from 
the pole configurations and will discuss later the result of the 
limit. For the time being we stress that the displacement has to be 
the {\em same} for all rapidities. Any different choice would introduce
an unjustified asymmetry among the particles.

Notice that, as a combined effect of the delta functions associated with
the disconnected parts, the $S$-matrix appears explicitely in the 
expressions for the $f_n^\Phi$ only through the constant factor $S(0)$. 
It follows from Eq.\,(\ref{unitarity}) that 
\EQ
S(0)=\pm 1\,\,.
\EN
Eq.\,(\ref{exchange}) shows that $S(0)$ determines whether the statistics 
of the particles is bosonic or fermionic. This means that the expansion of 
$\langle\Phi\rangle_R$ in the generic integrable case is formally the same
than in the free bosonic and fermionic cases,
all the difference being encoded in the explicit expressions of the form 
factors.

The sum of the $f_n^\Phi$ we determined gives
\bea
f^\Phi(R) &=& Z\,\{{\cal F}^\Phi_0+\frac{1}{2\pi}\,\int d\th_1\,e^{-E_1R}\,
[1+S(0)e^{-E_1R}+e^{-2E_1R}]\,{\cal F}^\Phi_2 \nonumber \\
&+& \frac{1}{2}\,\frac{1}{(2\pi)^2}\,\int d\th_1d\th_2\,e^{-E_2R}\,
[1+(e^{-MR\cosh\th_1}+e^{-MR\cosh\th_2})S(0)]\,{\cal F}^\Phi_4(\th_1,\th_2)
\nonumber \\
&+& \frac{1}{3!}\,\frac{1}{(2\pi)^3}\int d\th_1d\th_2d\th_3\,e^{-E_3R}\,
{\cal F}^\Phi_6(\th_1,\th_2,\th_3)+O(e^{-4MR})\}\,,
\eea
where all the singular delta function terms factorise into the quantity
\bea
Z &=& 1+\int d\th_1\,e^{-E_1R}\delta(0)+\frac{1}{2}\,\int d\th_1d\th_2\,
e^{-E_2R}\,[S(0)\delta^2(\th_1-\th_2)+\delta^2(0)]\nonumber \\
&+& \frac{1}{3!}\,\int d\th_1d\th_2d\th_3\,e^{-E_3R}\,
[2\delta(\th_1-\th_2)\delta(\th_1-\th_3)\delta(\th_2-\th_3)+
3S(0)\delta(0)\delta^2(\th_1-\th_2)+\delta^3(0)] \nonumber \\
&+& O(e^{-4MR})\,\,.
\eea
Since ${\cal F}^I_{2n}=\delta_{n,0}$, we have $f^I=Z$, so that the 
singular disconnected parts completely cancel out in the ratio 
(\ref{ratio}) without the need of any specific regularisation.

One can continue the computation and work out the next contributions
with $n>3$ to convince himself that the complete result is 
\EQ
\langle\Phi\rangle_R=\sum_{n=0}^\infty\,\frac{1}{n!}\,\frac{1}{(2\pi)^n}\,
\int\left[\prod_{i=1}^nd\th_i\,g(\th_i,R)\,e^{-MR\cosh\th_i}\right]\,
{\cal F}^\Phi_{2n}(\th_1,\ldots,\th_n)\,,
\label{final}
\EN
with
\EQ
g(\th,R)=\frac{1}{1-S(0)e^{-MR\cosh\th}}\,\,.
\label{g}
\EN
This expression provides the explicit form factor expansion for the
one-point function. It only remains to check that the specific form factor
configuration (\ref{parallel}) is not singular due to the annihilation poles.
For this purpose, take the form factor $F^\Phi_{0,2n}(|\th_1,\ldots,\th_{2n})$ 
and bring the first two particles on the singular configuration 
$\th_1\rightarrow\th_2+i\pi$. This pinching produces a pole with the residue 
given by Eq.\,(\ref{res}). We then repeat the operation on the second pair of 
particles by taking $\th_3\rightarrow\th_4+i\pi$, and so on with all the 
$n$ pairs.
This procedure produces $n$ poles. The numerator will contain the product
\EQ
\prod_{k=1}^{n}\left[1-\prod_{i=2k+1}^{2n}S(\th_i-\th_{2k})\right]\,,
\label{prodotto}
\EN
where each string of scattering amplitudes inside the parenthesis is made
of factors $S(\th_{2j}+i\pi-\th_{2k})S(\th_{2j}-\th_{2k})$. It follows
from (\ref{crossing}) and (\ref{unitarity}) that such factors are equal
to $1$, so that (\ref{prodotto}) gives $n$ zeros which cancel the poles
and leave a finite result. We made these considerations on a form factor
which differs from (\ref{parallel}) only for the ordering of the particles,
namely (recall Eq.\,(\ref{exchange})) for an overall phase which does not
affect the final conclusion.

For later convenience, we write down
more explicitely the first few terms in the low temperature expansion.
They read 
\bea
\langle\Phi\rangle_R &=& {\cal F}^\Phi_0+\frac{1}{\pi}[K_0(r)+S(0)K_0(2r)]\,
{\cal F}^\Phi_2+\frac{1}{\pi^2}\int_0^\infty d\th\,K_0(2r\cosh\th)\,
{\cal F}^\Phi_4(2\th,0) \nonumber \\
&+& O(e^{-3r})\,, 
\label{lt}
\eea
with 
\EQ
r\equiv MR\,\,.
\EN

\vspace{.3cm}
For the specific case in which the operator $\Phi(x)$ coincides with the 
trace $\Theta(x)$ of the stress-energy tensor, an alternative and effective
way of computing the one-point function is provided by the thermodynamic
Bethe ansatz approach \cite{TBA}. It is then interesting to make a check of 
the agreement of the two methods in this case. The one-point function of 
the trace is related to the ground state energy on the cylinder (without
the bulk term) $E(r)$ as 
\EQ
\langle\Theta\rangle_R=\langle\Theta\rangle_{R=\infty}+2\pi\,\frac{M}{r}\,
\frac{d}{dr}[rE(r)]\,.
\label{diff}
\EN
The TBA allows the determination of $E(r)$ from the knowledge of the 
$S$-matrix in the form 
\EQ
E(r)=-M\int\frac{d\th}{2\pi}\,L(\th)\cosh\th \,,
\label{gs}
\EN
\EQ
L(\th)\equiv -S(0)\ln (1-S(0)e^{-\varepsilon(\th)})\,,
\EN
where the `pseudoenergy' $\varepsilon(\th)$ is determined by the 
integral equation
\EQ
\varepsilon(\th)=r\cosh\th-\frac{1}{2\pi}\int d\th'\,\varphi(\th-\th')
L(\th')\,,
\label{integral}
\EN
\EQ
\varphi(\th)\equiv -i\frac{d}{d\th}\ln S(\th)\,.
\EN
The pseudoenergy $\varepsilon(\th)$ tends to $r\cosh\th$ in the zero 
temperature limit. Iterating (\ref{integral}) once with this initial condition 
and substituting into Eq.\,(\ref{gs})
produces the result\footnote{In the computation one uses the fact that 
$\varphi(\th)$ is an even function.}
\EQ
\frac{E(r)}{M}=-\frac{1}{\pi}\,[K_1(r)+S(0)\frac{1}{2}K_1(2r)]-\frac{2}
{\pi^2}\,\int_0^\infty d\th\,K_1(2r\cosh\th)\varphi(2\th)\cosh\th+O(e^{-3r})\,.
\EN
We can now evaluate $\langle\Theta\rangle_R$ to this order from (\ref{diff})
using the identity $\frac{d}{zdz}[z K_1(z)]=-K_0(z)$.
The final result exactly coincides with (\ref{lt}) because it was shown in
Ref.\,\cite{nonint} (Eqs.\,(3.23), (3.24)) on completely general and 
independent basis that\footnote{It has been checked
for many models that only the $\epsilon$-prescription (\ref{parallel}) 
reproduces the general result (\ref{f4}).}
\bea
&& {\cal F}^\Theta_2=2\pi M^2\,,\label{f2}\\
&& {\cal F}^\Theta_4(\th_1,\th_2)=8\pi M^2\,\varphi(\th_1-\th_2)
\cosh^2\frac{\th_1-\th_2}{2}\,.\label{f4}
\eea

We also recall here that nearby a fixed point the ground state energy 
behaves as \cite{Cardy} $E(R)\simeq -\pi C_{eff}/6R$, where
\EQ
C_{eff}=C-\frac{X_{min}}{12}\,\,,
\EN
is given in terms of the central charge $C$ and smallest scaling dimension
$X_{min}$ at the conformal point ($X_{min}=0$ in a `unitary' theory).
Integration of Eq.\,(\ref{diff}) then gives
\EQ
C_{eff}^{UV}-C_{eff}^{IR}=\frac{3}{\pi^2M^2}\int_0^\infty dr\,r
(\langle\Theta\rangle_R-\langle\Theta\rangle_{R=\infty})\,,
\label{sr}
\EN
for the total variation along the whole renormalisation group flow
($C^{IR}_{eff}=0$ in a massive theory). It is
easy to check that inserting in this formula Eq.\,(\ref{final}) with 
(\ref{f2}) as the only non-zero contribution for $n>0$ gives the expected
results $1$ for the free boson and $1/2$ for the free fermion. One can
also check (in the free as in the interacting case) that the approximation
of the r.h.s. of (\ref{sr}) through the first low temperature terms
(\ref{lt}) produces a poor result, indicating that the convergence to the 
intermediate and high temperature regimes is not very rapid. This fact
can be contrasted with the impressively fast convergence of the form 
factor expansion (as a function of the distance) for the correlation 
functions in integrable field theory at zero temperature. 
This different behaviour is hardly surprising in view of the different 
nature of the two expansions.

Since the distribution (\ref{g}) becomes singular as $R\rightarrow 0$ in
the bosonic case but not in the fermionic case, the high temperature 
behaviour is very sensitive to the statistics. For the quadratic operators
in the free cases one has
\EQ
[\langle\Theta\rangle_R-\langle\Theta\rangle_{R=\infty}]_{free}=
2M^2\sum_{k=1}^\infty S^{k-1}K_0(kr)
\simeq\left\{\matrix{\pi M^2/r\,,\hspace{.8cm}S=1\cr\cr
-M^2\ln r\,,\hspace{.8cm}S=-1\,.\cr}\right.
\label{freetrace}
\EN
For the vertex operators $V_a=e^{a\varphi}$ with scaling dimension 
$X_a=-a^2/4\pi$, in the free theory one has ${\cal F}^{V_a}_{2n}=a^{2n}/2^n$, 
so that (\ref{final}) gives
\EQ
\langle e^{a\varphi}\rangle_R^{free}\simeq e^{a^2/4r}\,\,,\hspace{1cm}
R\rightarrow 0\,.
\label{essential}
\EN

In the bosonic case $S(0)=1$, it follows from the $1/r$ 
behaviour of the function (\ref{g}) that the $n$-particle contribution
to $\langle\Phi\rangle_R$ behaves as $1/r^n$ at high temperature. Up to 
unlikely cancellations of infinities, this means that the integral
in the sum rule (\ref{sr}) diverges if $\langle\Theta\rangle_R$ receives
contributions with $n>1$, namely if the theory is interacting. Therefore,
one is led to the conclusion that in two dimensions the bosonic statistics
applies only to the free case\footnote{See Ref.\,\cite{MS} for an example 
of the troubles one encounters when looking for a counterexample.} $S(\th)=1$. 
An illustration of this fact is given by the sinh-Gordon model, namely the 
theory of a scalar field self-interacting trough the potential 
$\mu\cosh g\varphi$. Its exact scattering amplitude 
$S(\th)=(\sinh\th-i\sin\pi B)/
(\sinh\th+i\sin\pi B)$, with $B=g^2/(8\pi+g^2)$, shows that $S(0)$ becomes
$-1$ as soon as $g$ is taken different from zero. One can expect that 
the one-point function $\langle e^{a\varphi}\rangle_R$ will 
behave in the high temperature limit as a power law with an exponent 
that diverges as $g\rightarrow 0$ in order to reproduce the essential 
singularity (\ref{essential}).

\vspace{.3cm}
We said that the operator $\Phi(x)$ entering (\ref{ratio}) has to be 
local with respect to the excitations of the zero temperature theory.
We now clarify the origin of this requirement.
The general situation can be illustrated through the example 
of the Ising field theory defined at $R=\infty$ by the action
\EQ
{\cal A}=A_{CFT}+\tau\int d^2x\,\varepsilon(x)+h\int d^2x\,\sigma(x)\,,
\label{ising}
\EN
describing the perturbation of the Ising conformal point by the two 
relevant operators of the theory, the energy $\varepsilon(x)$ and the 
spin $\sigma(x)$. 

Consider first the case of vanishing magnetic field, $h=0$, in which
(\ref{ising}) describes a disordered phase ($\tau>0$) and an ordered 
phase ($\tau<0$) dual to each other and both corresponding to a free 
fermionic theory ($S=-1$). What changes in the scattering description
of the two phases is the nature of the particles, which are ordinary 
excitations over the unique vacuum at $\tau>0$, and kinks interpolating
between the two degenerate vacua of the spontaneously broken phase 
at $\tau<0$. The perturbing operator $\varepsilon$ is proportional to 
the trace of the stress-energy tensor ($\Theta\sim\tau\varepsilon$) and 
is local with respect to both kind of excitations. Its finite temperature 
one-point function is given by (\ref{freetrace}). The form factors of 
the spin operator are known in both phases \cite{BK,YZ}.
At $\tau>0$, they are non-zero only on states with an odd number of 
particles so that $\langle\sigma\rangle_R$ vanishes, as it should by
spin reversal symmetry. In the broken symmetry phase the spin
couples to the states with an even number of kinks, and one could expect
a non-trivial result from Eq.\,(\ref{final}). The spin, however, is not
local with respect to the kinks, and the residue (\ref{res}) on the 
annihilation poles gets modified by a constant phase factor (a minus sign
in this specific case) multiplying the product of scattering amplitudes 
\cite{YZ}. A similar modification is induced in the factor (\ref{prodotto}) 
which no longer cancels the annihilation poles of the matrix elements
(\ref{parallel}), so that the expression (\ref{final}) becomes badly
divergent for $R<\infty$. The conclusion is that $\langle\sigma\rangle_R$
is not defined at $\tau<0$. Since a finite value of $\langle\sigma\rangle_R$
would imply the existence of spontaneous symmetry breaking on the cylinder,
this result illustrates the absence of phase transitions in systems with a 
single infinite dimension. 

The same mechanism applies to the generic case. In two (infinite) 
dimensions, only a discrete symmetry can be broken spontaneously. The 
elementary excitations in the broken phase are kinks interpolating among
the discrete vacua. The order parameter is non-local with respect to them,
and its expectation values on the asympotic states are incurably
divergent.

Let us conclude this discussion of the Ising field theory at finite 
temperature by considering the other integrable direction of the action 
(\ref{ising}), namely the case $h\neq 0$, $\tau=0$. The spectrum of 
the theory \cite{Taniguchi} contains now three particles below the 
lowest two-particle threshold with masses $M_1$, 
\bea
&& M_2=(1.6180339887..)\,M_1\,,\nonumber \\
&& M_3=(1.9890437907..)\,M_1\,.\nonumber 
\eea
Hence, the first terms of the low-temperature expansion for the one-point 
functions are
\EQ
\frac{\langle\Phi\rangle_R}{\,\,\,\,\langle\Phi\rangle_{R=\infty}}=1+
\frac{1}{\pi}\sum_{i=1}^3 A_i^\Phi\,K_0(M_iR)+O(e^{-2M_1R})\,,
\label{universal}
\EN
where the ratios\footnote{Here $a_i(\th)$ denotes a particle with mass $M_i$ 
and rapidity $\th$.}
\EQ
A_i^\Phi=\frac{\langle 0|\Phi(0)|a_i(i\pi),a_i(0)\rangle}
{\langle\Phi\rangle_{R=\infty}}\,,
\EN
are {\em universal} and can be extracted from the work of 
Refs.\,\cite{immf,DS} 
on the form factors of the Ising model in a magnetic field. We give their 
values in Table\,1.


\vspace{.5cm}
\begin{center}
\begin{tabular}{|c||c|c|}\hline
$\Phi$      & $\sigma$ & $\varepsilon$ \\ \hline
$ A^\Phi_1$ & $-8.0999744..$ & $-17.893304..$      \\
$ A^\Phi_2$ & $-21.206008..$ & $-24.946727..$      \\
$ A^\Phi_3$ & $-32.045891..$ & $-53.679951..$      \\ \hline
\end{tabular}
\end{center}
{\bf Table 1.} The universal amplitudes entering the expansion
(\ref{universal}) for the Ising model in a magnetic field.

\vspace{.5cm}
\noindent
Notice that in this case the trace of the stress-energy tensor is 
proportional to the spin operator. Hence, the amplitudes $A^\sigma_i$
are $2\pi M_i^2$ divided by $\langle\sigma\rangle_{R=\infty}$. The latter
quantity is also known from the TBA \cite{Fateev}. The amplitudes
$A_i^\varepsilon$, instead, can only be obtained through a form factor 
computation \cite{DS} in which the `cluster' property explained in 
Ref.\,\cite{DSC} plays an essential role.

\vspace{.3cm}
In conclusion, the form factor approach provides a systematic and explicit 
low temperature expansion for the one-point functions in integrable field 
theory at finite temperature. Form factors have already been computed
for many integrable theories and it is highly desirable to
have numerical data to compare with the theoretical predictions. 
It also seems natural to expect that the study of the two-point 
functions should be approachable along the same lines with the goal of 
obtaining a low temperature, large distance expansion. 
This problem is beyond the scope of this note and will be considered
elsewhere.

\vspace{.5cm}
\noindent
{\bf Acknowledgments:} I thank G. Mussardo for interesting discussions.


\begin{thebibliography}{99}
\bibitem{LLSS} A. LeClair, F. Lesage, S. Sachdev and H. Saleur,
{\em Nucl. Phys.} {\bf B 482} (1996), 579.
\bibitem{LM} A. LeClair and G. Mussardo, {\em Nucl. Phys.} {\bf B 552}
(1999), 624.
\bibitem{Saleur} H. Saleur, {\em Nucl. Phys.} {\bf B 567} (2000), 602.
\bibitem{Lukyanov} S. Lukyanov, Finite temperature expectation
values of local fields in the sinh-Gordon model, hep-th/0005027.
\bibitem{ZZ} A.B. Zamolodchikov and Al.B. Zamolodchikov, {\em Ann. Phys.} 
{\bf 120} (1979), 253.
\bibitem{Karowski} M. Karowski and P. Weisz, {\em Nucl. Phys.} {\bf B 139}
(1978), 455.
\bibitem{Smirnov} F.A. Smirnov, Form Factors in Completely Integrable
Models of Quantum Field Theories (World Scientific) 1992.
\bibitem{nonint} G. Delfino, G. Mussardo and P. Simonetti, {\em Nucl. Phys.}
{\bf B 473} (1996), 469.
\bibitem{TBA} Al.B. Zamolodchikov, {\em Nucl. Phys.} {\bf B 342} (1990),
695.
\bibitem{Cardy} J.L. Cardy, {\em J. Phys.} {\bf A 16} (1984), L385.
\bibitem{MS} G. Mussardo and P. Simon, {\em Nucl. Phys.} {\bf B 578} 
(2000), 527.
\bibitem{BK} B. Berg, M. Karowski and P. Weisz, {\em Phys. Rev.}
{\bf D 19} (1979), 2477.
\bibitem{YZ} V.P. Yurov and Al.B. Zamolodchikov, {\em Int. J. Mod. Phys.}
{\bf A6} (1991), 3419.
\bibitem{Taniguchi} A.B. Zamolodchikov, {\em Adv. Stud. Pure Math.}
{\bf 19} (1989), 641.
\bibitem{immf} G. Delfino and G. Mussardo, {\em Nucl. Phys.} {\bf B 455} 
(1995), 724.
\bibitem{DS} G. Delfino and P. Simonetti, {\em Phys. Lett.} {\bf B 383}
(1996), 450.
\bibitem{Fateev} V.A. Fateev, {\em Phys. Lett.} {\bf B 324} (1994), 45.
\bibitem{DSC} G. Delfino, P. Simonetti and J.L. Cardy, {\em Phys. Lett.} 
{\bf B 387} (1996), 327.


\end{thebibliography}
\end{document}